\documentclass{pas}
\usepackage{aas-macros}
\usepackage{amsmath}
\usepackage{natbib}
\usepackage{multirow}
\usepackage{graphicx}
\usepackage{appendix}
\usepackage{etex}
\usepackage{gensymb}
\usepackage{amsmath,siunitx}
\usepackage [english]{babel}
\usepackage [autostyle, english = american]{csquotes}
\MakeOuterQuote{"}

\begin{document}

\lefttitle{Resolved K-S relation in nearby galaxies using UVIT and ALMA observations}
\righttitle{Sruthi et al. 2025}

\jnlPage{1}{17}
\jnlDoiYr{2025}
\doival{10.1017/pasa.xxxx.xx}

\articletitt{Research Paper}

\title{Probing the resolved K-S relation in nearby galaxies: Insights from UVIT and ALMA observations}

\author{\sn{K} \gn{Sruthi}, Sreeja S Kartha, Blesson Mathew, K. Ujjwal, Akhil Krishna R and Shankar Ray}

\affil{Centre of Excellence in Astronomy and Astrophysics, Department of Physics and Electronics, CHRIST (Deemed to be University), Bangalore, India}

\corresp{Sruthi K, Email: sruthi.k@christuniversity.in}

\history{(Received xx xx 2024; revised xx xx 2024; accepted xx xx 2024)}

\begin{abstract}
This study examines the resolved Kennicutt-Schmidt (rK-S) relation, defined as the connection between the star formation rate surface density ($\Sigma_{SFR}$) and the molecular gas mass surface density ($\Sigma_{H_2}$) in the high-density central regions of three nearby barred spiral galaxies hosting AGN: NGC 1365, NGC 1433, and NGC 1566. Utilising high-resolution archival data from AstroSat/UVIT for UV imaging and Atacama Large Millimetere/submillimetre Array (ALMA) for CO(2-1) molecular gas mapping, we explore recent star formation and gas distribution with a spatial resolution of $\sim$ 120-132 pc. Our findings reveal a sublinear rK-S law, with slopes ranging from $\sim$ 0.17 to $\sim$ 0.71. Notably, NGC 1566 exhibits a robust rK-S relation consistent with previous studies, while NGC 1365 and NGC 1433 exhibit weaker correlations. These differences are likely due to the smaller number of identified star-forming regions in these galaxies compared to NGC 1566, as well as the central molecular gas concentrations and varying star formation activity in their bars and nuclear regions. These results also support the idea that the rK-S relation deviates from linearity in extreme environments, such as starburst galaxies and galactic centres. Additionally, we find a generally low median star formation efficiency (SFE) within the bars of these galaxies, suggesting that while bars may drive nuclear starbursts and contribute to bulge growth, they do not significantly increase SFE. Furthermore, a negative correlation between SFE and $\Sigma_{H_2}$ is observed across the sample, both within and outside the bar regions, suggesting that higher $\Sigma_{H_2}$ may lead to lower SFE in the central regions of these galaxies. Our findings highlight that $\Sigma_{H_2}$ plays a primary role in shaping the observed trends in SFE, rather than the presence of a bar itself.
\end{abstract}


\maketitle
\section{Introduction}

Stars form through the gravitational collapse of cold, dense gas within giant molecular clouds (hereafter GMCs), which are characterised by supersonic turbulence, magnetism, and gravitational forces \citep{2012ARA&A..50..531K,2014MNRAS.439.3420M}. The rate of star formation is therefore governed by the properties of GMCs \citep{2022ApJ...927....9P}. Understanding the relationship between gas and star formation in galaxies is essential for studying galaxy evolution, as it determines the efficiency of gas conversion into stars and provides critical insights for models and simulations \citep{2005ApJ...630..250K,2003MNRAS.339..289S}. However, the mechanisms controlling gas-to-star conversion across different galactic environments, such as the nuclear region, bar region, and spiral arms, remain poorly understood. This is primarily due to the complex interplay of processes within the interstellar medium (ISM), which involve geometrical, dynamical, and chemical properties \citep{2021ApJ...913..139G}. Additionally, feedback from young stellar objects \citep[e.g.][]{2016ApJ...822...49J}, supernovae \citep[e.g.][]{2020MNRAS.493.4700L}, and active galactic nuclei \citep[hereafter AGN;][]{2015A&A...577A.135C,2021ApJ...913..139G} further complicates our understanding.

The relationship between the star formation rate (hereafter SFR) and the gas content in galaxies was initially proposed by \cite{1959ApJ...129..243S,1963ApJ...137..758S}, suggesting an equation of the form, $\Sigma_{\text{SFR}} = A \Sigma_{\text{gas}}^{N}$.  Here, $\Sigma_{\text{SFR}}$ and $\Sigma_{\text{gas}}$ denote the surface densities of SFR and gas, respectively. The parameter A represents the normalisation constant, signifying the efficiency of processes governing the conversion of gas into stars, while \emph{N} represents the power-law index. The gas component can comprise atomic (HI), molecular (H$_{\text{2}}$), or a combination of both (HI+H$_{\text{2}}$). \cite{1959ApJ...129..243S} derived a power-law index, approximately $N \approx 2$, in the above Equation by comparing the atomic gas density with the number of representative young stellar objects in the solar neighbourhood. \cite{1978A&A....68....1G} further confirmed values of $N$ ranging approximately from 1.5 to 2. Subsequently, \cite{1998ARA&A..36..189K,1998ApJ...498..541K} discovered that for disk-averaged surface densities, both normal star-forming and starburst galaxies adhere to this Equation with a power-law index of approximately $N \approx 1.4$ for total (HI+H$_{\text{2}}$) gas. This correlation between gas and SFR surface densities is widely known as the Kennicutt-Schmidt relation (hereafter the K-S relation). Current models of star formation suggest a connection between star formation, feedback, and gas properties at the scale of individual GMCs, which can be inferred using high resolution observations \citep{2021ApJS..257...43L}. As observational evidence supporting this, recent studies \citep{2008AJ....136.2782L,2011AJ....142...37S,2019MNRAS.483.5135W,2019MNRAS.488.1926D} have demonstrated that this relationship remains valid even at kiloparsec and sub-kiloparsec spatial scales, approaching the intrinsic scale of star formation, namely the size of GMCs \citep[$\sim$10-150 pc, e.g.,][]{1987ApJ...319..730S}. However, \citet{2014MNRAS.439.3239K} demonstrated that star formation scaling relations exhibit significant scatter or even change form below certain spatial scales. They proposed an uncertainty principle \citep{1927ZPhy...43..172H} for star formation, which provides a framework for predicting and interpreting the breakdown of galactic star formation relations at smaller spatial scales.
Additionally, there is an ongoing debate on the slope of these scaling relationships, and it is acknowledged that their values are influenced by the specific methodology used for their calculation \citep{2021A&A...650A.134P}. The considerable variation in the value of \emph{N} could originate intrinsically, indicating the existence of different star formation "laws" that potentially hold valuable astrophysical insights \citep{2015A&A...577A.135C}. Alternatively, this variability may be partly or entirely attributed to factors such as the chosen physical scale \citep{2010ApJ...722.1699S}, the selection of molecular gas tracer \citep{2008ApJ...684..996N}, the type of galaxy under investigation, data sampling, and the fitting methodology employed \citep{2013MNRAS.430..288S}. The previous literature has found resolved K-S relation (hereafter rK-S) that range from highly sublinear to quadratic \citep[e.g.,][]{2008AJ....136.2846B,2010ApJ...714L.118D,2012MNRAS.421.3127N,2014MNRAS.442.2208S, 2019ApJ...884L..33L,2019ApJ...872...63L, 2020MNRAS.492.6027E, 2021MNRAS.503.1615S, 2021A&A...650A.134P, 2022A&A...663A..61P}. Furthermore, the dispersion observed in the rK-S relation has been explained due to individual star-forming regions that undergo distinct evolutionary life cycles \citep{2010ApJ...722.1699S}. Moreover, its variation with spatial scale offers insights into the timescales associated with the star formation cycle \citep{2014MNRAS.439.3239K}. The correlation between $\Sigma_{SFR}$ and $\Sigma_{H_2}$ shows substantial scatter when the spatial resolution is adequate to distinguish the individual elements of the surface densities, such as GMCs and star-forming regions \citep{2022ApJ...927....9P}. However, the observation of this scaling relationship at the resolved scale indicates that its global counterpart arises from the local mechanisms that govern star formation. Examining these relationships and their associated variability provides valuable insights into the local physical processes that regulate the formation of stars.

A critical factor in investigating the rK-S relation is the selection of tracers for SFR and H$_{\text{2}}$. 
Various SFR tracers, including H$\alpha$ emission, far-ultraviolet (FUV) emission, and 24 $\mu$m mid-infrared emission, reflect different timescales in a galaxy's star formation history \citep{2015A&A...577A.135C}. 
The two most commonly used tracers of current star formation are H$\alpha$ emission and UV continuum emission. H$\alpha$ is produced by the recombination of ionised hydrogen \citep{2024ApJ...976...90T}. In star-forming regions, the gas is ionised by young, massive stars, specifically O-type stars with lifespans of $\leq 7$ Myr and masses of $M_* \geq 20 \, M_\odot$, and B-type stars with lifespans of $\leq 320$ Myr \citep{2009ApJ...695..765M}, found in OB associations. Due to the short lifespan of O-type stars, H$\alpha$ emission typically lasts no more than 10 Myr, making it a tracer of the most recent SF \citep{2024ApJ...976...90T}.  In contrast, UV emission originates as direct light from newly formed but less massive stars, including O-type stars with masses of $M_* \geq 3 \, M_\odot$ \citep{2009ApJ...695..765M}. These stars have longer lifespans, ranging from 7 Myr to up to 200 Myr, enabling UV emission to trace not only the most recent SF but also phases of the SF history older than the H$\alpha$-emitting phase \citep{2024ApJ...976...90T}.
Therefore, in general, FUV is considered as an excellent tracer of recent star formation in galaxies \citep{1998ARA&A..36..189K}. Regions undergoing active star formation, marked by the presence of young, hot, massive, and luminous O, B, A stars that emit substantial amounts of UV radiation. Consequently, these star-forming regions appear bright in the UV images. Hence, the ultraviolet continuum serves as a straightforward and direct indicator of recent star formation in galaxies, typically spanning a timeframe of $\sim$ 200 million years. 
Over the past decade, the Ultraviolet Imaging Telescope (UVIT), onboard AstroSat, has provided a new avenue for conducting high-resolution UV studies of galaxies in the nearby Universe \citep[e.g.,][]{2018A&A...614A.130G,2018MNRAS.479.4126G,2018AJ....156..109M,2021ApJ...914...54Y,2022MNRAS.516.2171U,2023ApJ...946...65D,2023MNRAS.522.1196R,2024A&A...681A...7R,2024A&A...684A..71U}. 
Consequently, UVIT has proven to be a valuable tool for detecting star-forming regions within nearby galaxies.

In barred spiral galaxies, star formation varies across different regions such as the spiral arms, interarms, and bar regions \citep[e.g.,][]{2010ApJ...725..534F,2013MNRAS.431.1587E, 2024arXiv240505364Q}. Optical and millimetre/submillimetre observations indicate concentration of young stars and molecular gas clouds in spiral arms, showing higher SFR densities and molecular gas densities \citep{2021ApJ...913..139G}. In barred galaxies hosting AGNs, star formation in nuclear regions can be influenced by outflows and radio jets from the central AGN. This influence may lead to either suppression \citep{2018NatAs...2..198H} or enhancement \citep{2019MNRAS.485.3409G} of star formation. The impact of AGN feedback on star formation remains a longstanding challenge, with recent studies reporting both negative and positive feedback \citep[e.g.][]{2018A&A...613L...9G,2019ApJ...881..147S,2022MNRAS.516.2300J,2024arXiv240618103N}. In this context, it is worthwhile to further explore the influence of bars and AGNs on the star formation properties of galaxies. This can be achieved by analysing the star formation and molecular gas properties using existing scaling relations, such as the rK-S relation.

The aim of this paper is to examine the rK-S relation within the inner areas ($<$ 22 kpc) of three nearby ($<$ 20 Mpc) barred spiral galaxies hosting AGN at a spatial scale from $\sim$ 120 pc to $\sim$ 130 pc by analysing the emission from individual star-forming regions identified across the available maps. We concentrate exclusively on the relationship between star formation rate (SFR) and the molecular component of gas, as the gas phase in the central regions of galaxies is primarily molecular in nature \citep{2020A&A...641A.151S}. Leveraging the superior spatial resolution of UVIT and the Atacama Large Millimetre/submillimetre Array (ALMA), we investigate star-forming regions and GMCs in the vicinity of AGNs in the sample galaxies. This approach allows us to gain deeper insights into the star formation process. In this study, we examine the rK-S relation across the three galaxies in our sample, assessing its robustness and applicability within their distinct environments.

This paper is organised as follows. In Section \ref{sec:sample}, we provide a detailed description of the sample. Section \ref{sec:data} outlines the data used, their processing, and the methodology employed to derive $\Sigma_{SFR}$ and $\Sigma_{H_2}$ from the original images. This is followed by the analysis and results in Section \ref{sec:results}. A detailed discussion of the results is presented in Section \ref{sec:discussion}. The conclusion of the study is described in Section \ref{sec:conclusion}.  Throughout this paper, a flat Universe cosmology with H$_0$ = 71 km s$^{-1}$ Mpc$^{-1}$ and $\Omega_M$ = 0.27 is adopted, as per \citet{2011ApJS..192...18K}.

\section{The Sample}
\label{sec:sample}

\begin{figure*}[htbp]
\centering
\includegraphics[width=1\linewidth]{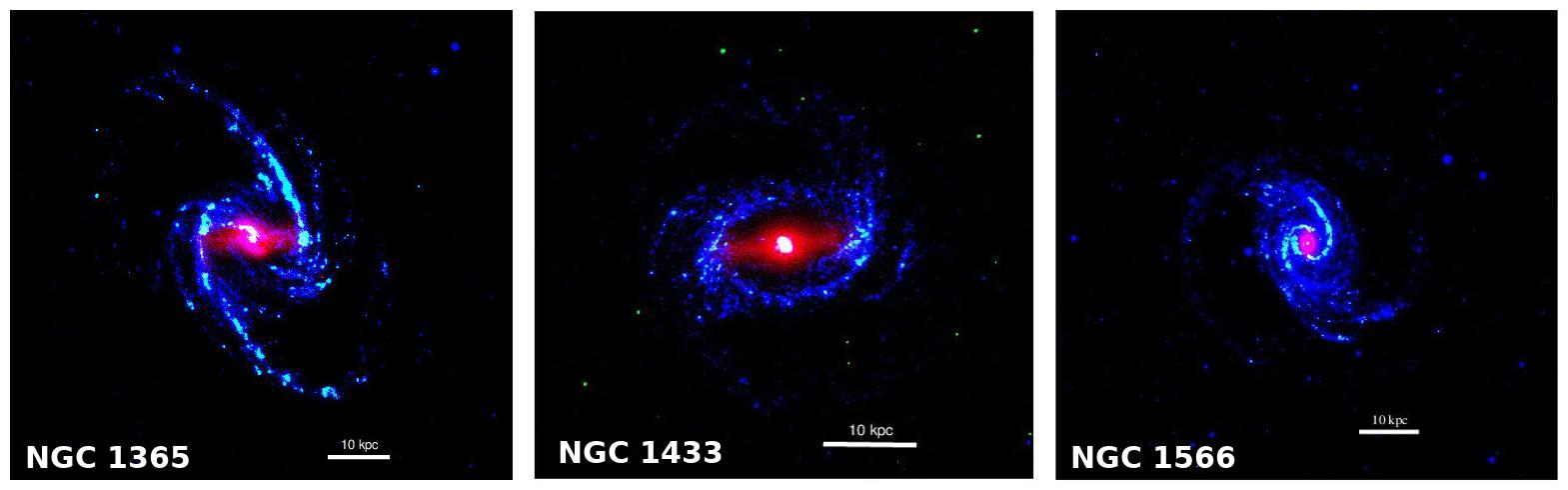}
\caption{The colour composite images of the sample galaxies are created using data from James Webb Space Telescope (JWST) F200W \citep[obtained from PHANGS-JWST archive;][]{2023ApJ...944L..17L}\protect\footnote{\url{https://www.canfar.net/storage/vault/list/phangs/RELEASES/PHANGS-JWST}}, UVIT FUV 148W/154W, and NUV 263M filters, with each filter shown in red, green, and blue colours, respectively. For all the images, North is up and East to the left.}

\label{fig:rgbimage}
\end{figure*}

To examine the relationship between star formation and molecular gas at a resolved level requires the proximity of the galaxies. Therefore, we have selected three nearby barred spiral galaxies hosting AGN namely NGC 1365, NGC 1433, and NGC 1566, which are located within 20 Mpc, and have comparable stellar masses \citep[$\sim$ 10$^{10}$ M$_{\odot}$;][]{2021A&A...656A.133Q}. These galaxies were selected primarily due to the availability of both UVIT and ALMA observations. Additionally, we prioritized the presence of H$\alpha$ and H$\beta$ data from the Multi Unit Spectroscopic Explorer \citep[MUSE;][]{2010SPIE.7735E..08B} of the Very Large Telescope (VLT) to ensure accurate estimation of internal extinction. Furthermore, it is well established that galaxies with higher stellar masses tend to exhibit greater global molecular gas and SFR surface densities \citep{2021A&A...656A.133Q}. The following subsections offer a concise overview of these galaxies, along with a listing of the sample properties in Table \ref{table_1}. The colour composite images of these galaxies, created using JWST and UVIT filters, are shown in Figure \ref{fig:rgbimage}.

\subsection{NGC 1365}

NGC 1365, situated in the Fornax cluster with a fiducial redshift of \emph{z} = 0.005457 \citep{2023A&A...678A..43N}, stands out as one of the most notable nearby barred galaxies \citep{1997MNRAS.288..715R}. It resides at a distance of 19.57 $\pm$ 0.78 Mpc \citep{2021MNRAS.501.3621A} and hosts a Seyfert 1.5 type nucleus \citep{1997MNRAS.288..715R}. NGC 1365 is a face-on spiral galaxy characterised by a prominent bar structure \citep{2008A&A...484..341R}. The 11.9 $\mu$m imaging of this galaxy reveals an unresolved nucleus flanked by two point sources to the northeast, surrounded by faint, arm-like extended emission \citep{2008A&A...484..341R}. \citet{2005A&A...438..803G} identified these sources as young, massive star clusters embedded within the galaxy. Notably, it features a circumnuclear star-forming ring with a diameter of around 1.3 kpc \citep{2009A&A...505..589G}.  Recent studies \citep[e.g.][]{2021ApJ...913..139G,2023ApJ...944L..15S} have revealed that the inner regions (within $<$ 9$^{\prime\prime}$) of NGC 1365 exhibit higher SFRs compared to the outer regions, suggesting intensified star formation activities within the star-forming ring around the AGN. Furthermore,  \citet{2021ApJ...913..139G} have identified dense regions with molecular gas surface densities ($\Sigma_{H_2}$) exceeding 100 M$_{\odot}$ pc$^{-2}$, but with a significantly lower star formation efficiency (SFE). The molecular outflow at the centre of NGC 1365, likely driven by AGN nuclear winds or hot shocked gas \citep{2021ApJ...913..139G}, depletes molecular gas faster than it is consumed by star formation, indicating a negative feedback scenario. Additionally, high gas inflow rates may contribute, as the gas entering the ring does not immediately form stars but can accumulate, leading to some stochasticity in star formation \citep{2023ApJ...944L..15S}.

\subsection{NGC 1433}

NGC 1433 is a nearby active spiral galaxy located at a distance of 18.63 $\pm$ 1.86 Mpc. The galaxy is a member of the Dorado group and is considered as a prototypical example of ringed barred spiral galaxies \citep{1996ApJ...460..665R}. NGC 1433 has a fiducial redshift of \emph{z} = 0.003589 \citep{2011ApJS..197...28P}. The galaxy features a prominent primary stellar bar, extending $\sim$ 4 kpc in the east-west direction \citep{2014A&A...567A.119S}. Near-infrared imaging has further revealed a nuclear bar within a nuclear ring, approximately 400 pc in radius, which is also a site of intense star formation \citep{2013A&A...558A.124C}. This double bar structure was previously observed by \citet{1986ApJS...61..631B} and \citet{1997A&AS..125..479J}. ALMA observations of molecular $^{12}$CO(3-2) emission have identified a highly redshifted component just south of the nucleus, indicating a possible outflow \citep{2013A&A...558A.124C}. This flow might be related to the buckling of the bar \citep{2017MNRAS.470L.122P} and the formation of a boxy/peanut bulge \citep{2023A&A...671A...8D}. Spitzer data also reveal the presence of a pseudo-bulge and prominent nuclear spirals within NGC 1433 \citep{2010ApJ...716..942F}. Furthermore, recent research by \citet{2023A&A...671A...8D}, using integral field spectroscopic data from the Time Inference with MUSE in Extragalactic Rings \citep[TIMER;][]{2019MNRAS.482..506G} survey, estimates that the bar in NGC 1433 is around 7.5 billion years old.

\subsection{NGC 1566}

NGC 1566, the brightest member of the Dorado group \citep[based on total H$\alpha$ luminosity;][]{2009ApJ...703.1672K} has a fiducial redshift of z = 0.005017 \citep{2014ApJ...783..106L}. It is often described as a grand-design spiral galaxy, showcasing an intricate structure characterised by two sets of bisymmetric spirals \citep{2017MNRAS.468..509G}. Although the distance to NGC 1566 is subject to considerable uncertainty in the literature, we adopt a distance of 17.69 $\pm$ 2 Mpc based on \citet{2021MNRAS.501.3621A}. Classified as a low-luminosity AGN \citep{2024arXiv240411656S}, NGC 1566 is typically identified as a Seyfert galaxy, although its precise classification between Seyfert 1 and 2 may vary between studies. NGC 1566 is classified as an SAB(rs)b galaxy, indicating an intermediate-type barred spiral galaxy with moderate apparent bar strength, featuring open, knotty arms, a small bulge, and an outer pseudo-ring composed of arms winding approximately 180$^{\circ}$ with respect to the bar ends \citep{2015ApJS..217...32B}. This galaxy serves as a typical example of a galaxy with bar-driven spiral density waves, as its strong spiral arms align with the region where bar driving is expected \citep{2017MNRAS.468..509G}. A recent study by \citet{2023A&A...673A.147P} suggests that the central regions of NGC 1566 appear to have undergone quenching, indicative of halted star formation activity.

\begin{table*}[h]
\centering
\caption{Details of the sample galaxies, with positional data sourced from NASA/IPAC Extragalactic Database (\url{https://ned.ipac.caltech.edu/}). Distance measurements are derived from \cite{2021MNRAS.501.3621A}, while galaxy inclinations are adopted from \cite{2020ApJ...897..122L}. Details regarding the ALMA data are described in \citet{2021ApJS..257...43L}.}
\label{table_1}
\begin{tabular}{llllllllll}
\hline
Name     & RA           & DEC           & Distance & Inclination & Obs.ID & \multicolumn{3}{c}{Details of UVIT Observation}         \\ 
         & (hh mm ss)   & (dd mm ss)    & (Mpc)    & (deg)       &    & & Exp. Time (s) & FUV Filter & Mean $\lambda$ ($\AA$) \\ 
         
         \hline
NGC 1365 & 03 33 36.371 & -36 08 25.447 & 19.57    & 55.4        & G07\_057 &      & 1162.61      & F148W  & 1481 \\ \hline
NGC 1433 & 03 42 01.550 & -47 13 19.500 & 18.63    & 28.6        & G07\_066 &     & 1976.71      & F154W  & 1541 \\ \hline
NGC 1566 & 04 20 00.420 & -54 56 16.100 & 17.69    & 29.5        & G06\_087 &      & 2940.39      & F148W  & 1481 \\ \hline
\end{tabular}
\newline

\end{table*}

\section{Data $\&$ Analysis}
\label{sec:data}
Our study focuses on determining the surface densities of SFR and molecular gas, $\Sigma_{SFR}$ and $\Sigma_{H_2}$, respectively, through the analysis of multiwavelength datasets. To estimate these densities, we utilised UVIT FUV emission images from the CaF2-1 (1481 $\AA$) and BaF2 (1541 $\AA$) filters for star formation and ALMA $^{12}$CO(2-1) line intensity maps for molecular gas. In this section, we provide an overview of these datasets and their application in our study.

\subsection{UVIT Data}
UVIT encompasses three distinctive spectral bands: the FUV band spanning 130-180 nm, the Near-UltraViolet (NUV) band ranging from 200 to 300 nm, and the Visible (VIS) band covering 320-550 nm. Notably, UVIT facilitates simultaneous observations across these bands, with the VIS channel instrumental in monitoring the satellite's positional stability during data acquisition. With a full circular field-of-view \citep[28$^\prime$ diameter;][]{2012SPIE.8443E..1NK}, UVIT provides spatial resolutions of about 1.4$^{\prime\prime}$ and 1.2$^{\prime\prime}$ for the FUV and NUV filters, respectively, with a pixel scale of $\sim$ 0.416$^{\prime\prime}$. The capabilities of UVIT make it especially capable of achieving exceptional spatial resolution in the UV region, surpassing its predecessor, the Galaxy Evolution Explorer (GALEX), and making it particularly well-suited for resolved scale studies in nearby galaxies.

\begin{figure*}[t]
\centering
\includegraphics[width=0.75\textwidth]{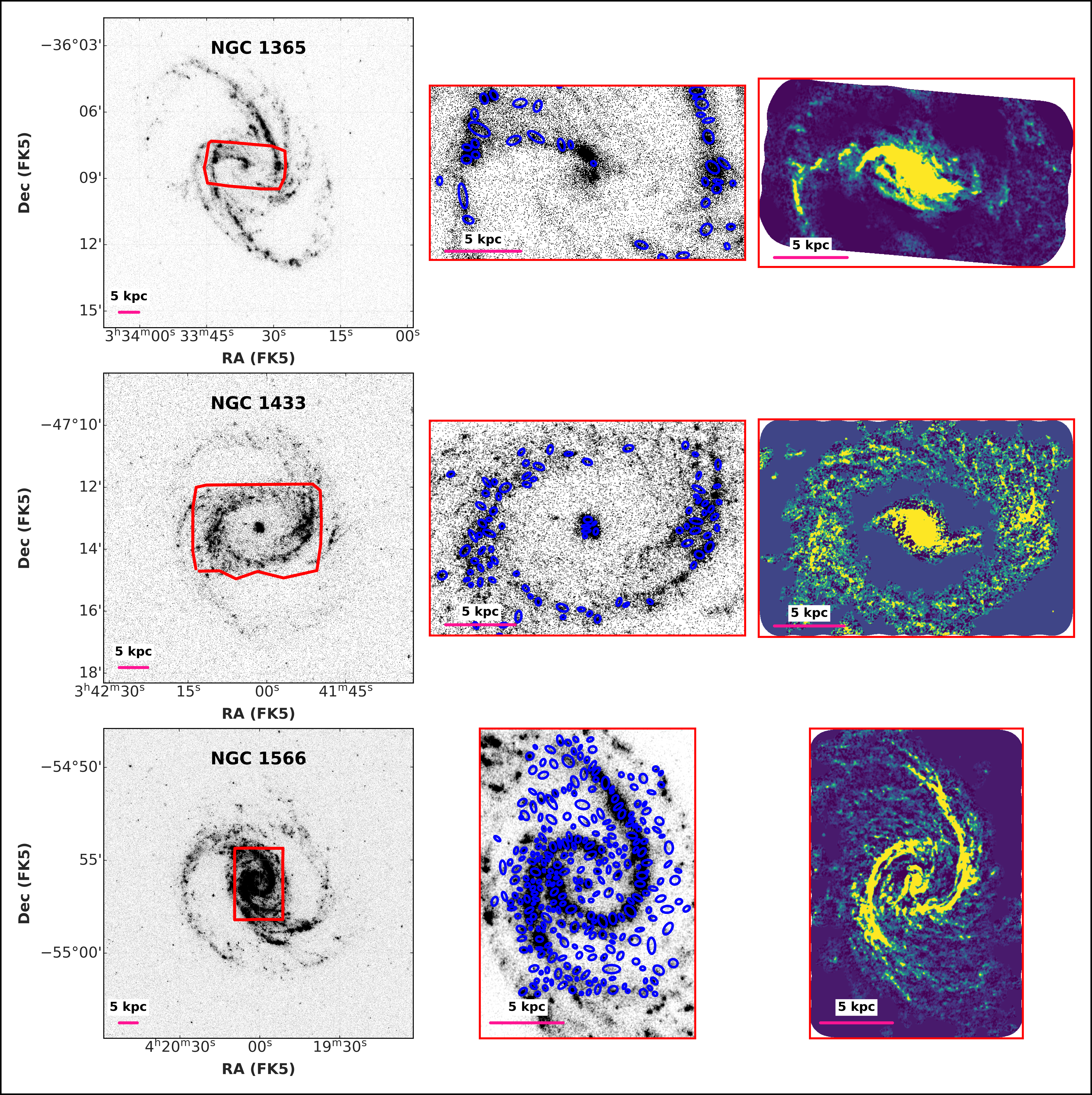}
\caption{The figure presents UVIT images of the galaxy sample in the left panel, with the ALMA-observed region outlined in red over the FUV images. The middle panel displays star-forming regions identified in UVIT within the ALMA FoV, overlaid on the FUV image. The right panel showcases the ALMA CO $^{12}$(2-1) image retrieved from the PHANGS archive. In all images, North is up, and East is to the left. AGN-contaminated FUV regions have been excluded from this visualisation.}
\label{fig_1}
\end{figure*}

Our investigation specifically relied on UVIT data archived at AstroSat Indian Space Science Data Centre \citep[ISSDC\footnote{\url{https://astrobrowse.issdc.gov.in/astro_archive/archive/Home.jsp}};][]{2012SPIE.8443E..1NK} for three galaxies: NGC 1365 (PI: C S Stalin), NGC 1433 (PI: Swarna), and NGC 1566 (PI: C S Stalin). F148W ($\lambda$$_{peak}$ $\sim$ 1481 $\AA$, $\Delta$$\lambda$ $\sim$ 500 $\AA$) images were used for NGC 1365 and NGC 1566, while F154W ($\lambda$$_{peak}$ $\sim$ 1541 $\AA$, $\Delta$$\lambda$ $\sim$ 380 $\AA$) was used for NGC 1433 due to the unavailability of F148W observation for this galaxy. We used the CCDLab software package \citep{2017PASP..129k5002P} to reduce the Level 1 data, with drift correction incorporated based on the VIS images \citep[see][]{2022MNRAS.516.2171U}. Subsequently, each image was flat-fielded, distortion corrected, and pattern noise reduction using the calibration files in CCDLab \citep{2017ExA....43...59G, 2017PASP..129k5002P, 2017AJ....154..128T, 2020AJ....159..158T}. Astrometric solutions were derived using CCDLab, using Gaia DR3 detections \citep{2023A&A...674A...1G} in the observed field \citep{2017PASP..129k5002P}. The UVIT FUV images, with a resolution of 1.4$^{\prime\prime}$, enable the resolution of star-forming regions up to $\sim$ 132 pc, 126 pc, and 120 pc in NGC 1365, NGC 1433, and NGC 1566, respectively. Detailed information regarding the UVIT observations, along with the position and distance data of the galaxies, is presented in Table \ref{table_1}. Figure \ref{fig_1} presents the UVIT images of the sample galaxies, with the ALMA observation field outlined in red colour.

\subsection{ALMA $^{12}$CO(2-1) Data}
The emission of CO lines serves as a marker for the widespread presence of molecular gas, which represents the cold, star-forming phase of the interstellar medium \citep{2021ApJS..257...43L}. In this investigation, we make use of the ALMA $^{12}$CO(2-1) moment0 maps of our sample galaxies, obtained from the Physics at High Angular resolution in Nearby GalaxieS (PHANGS) archive at the Canadian Astronomy Data Centre \citep[PHANGS-ALMA large program;][]{2021ApJS..257...43L,2021ApJS..255...19L}. These maps were derived from data cubes where a "broad" signal identification mask was applied. These broad masks encompass all sightlines where a signal is detected at any resolution, ensuring excellent completeness and broader coverage compared to strict maps \citep{2021ApJS..255...19L}.  The data consist of combined 12-m, 7-m, and total power array mapping with a typical native angular resolution of 1.38$^{\prime\prime}$ for NGC 1365, 1.10$^{\prime\prime}$ for NGC 1433, and 1.25$^{\prime\prime}$ for NGC 1566 \citep{2021ApJS..257...43L}. These values are broadly comparable to the spatial resolution achieved with UVIT, and in order to leverage UVIT’s higher resolution, the images were not convolved to a common resolution in this study. Additionally, we acquired error maps from the archive corresponding to the broad moment0 maps to calculate the errors associated with $\Sigma_{H_2}$. Further details on the reduction of ALMA images can be found in \citet{2021ApJS..257...43L}. The right panel of Figure \ref{fig_1} displays the ALMA images of the sample of galaxies.


\subsection{Identification of spatially resolved sites of star formation using UVIT}
\label{subsec:uvit}
To examine the resolved star formation characteristics within the sample, we utilised the ProFound package \citep{2018MNRAS.476.3137R}, an astronomical data-processing tool in the R programming language, to detect the most prominent regions in the UVIT/FUV images. By employing watershed deblending, ProFound identifies peak flux areas within the image and delineates the source segments. Subsequently, total photometry is estimated through iterative expansion (dilation) of the observed segments. Considering the spatial resolution of the UVIT FUV filter \citep[$\sim$ 1.4$^{\prime\prime}$;][]{2018A&A...614A.130G}, we established a criterion wherein the identified regions should encompass at least six pixels, equivalent to the minimum number required to cover a circle with a diameter equal to the FUV filter's spatial resolution. A $\emph{skycut}$ of 3 which corresponds to a detection threshold of 3$\sigma$ \citep {2018MNRAS.476.3137R} was applied to identify star-forming regions. Further details on source identification and background estimation can be found in \citet{2018MNRAS.476.3137R},\citet{2022MNRAS.516.2171U} and \citet{2024A&A...681A...7R}. Specifically, we identified 551 bright FUV regions in NGC 1365, 603 in NGC 1433, and 1801 in NGC 1566. This analysis yields fundamental information regarding the regions, including their positions, flux, and extent.  Subsequently, we calculated the magnitude of each identified region. Only star-forming regions with a photometric error of $\leq$ 0.1 are chosen for subsequent analysis, as outlined in \citet{2021ApJ...909..203M}. Additionally, any foreground stars are eliminated using $\emph{Gaia}$ DR3 catalogue \citep{2023A&A...674A...1G}. Consequently, the number of regions is reduced to 147 in NGC 1365, 108 in NGC 1433, and 691 in NGC 1566.


To account for AGN contamination in our analysis, we utilised the Baldwin, Phillips $\&$ Terlevich (BPT) diagrams \citep{1981PASP...93....5B},  specifically the [NII], [SII], and [OI] versions and adopted the corresponding AGN flags provided by \citet{2023MNRAS.520.4902G}, which identify AGN-ionised regions in these galaxies. A region was classified as AGN-affected if all three flags had a value of 3 (see \citet{2023MNRAS.520.4902G}). To assess the spatial overlap between FUV star-forming regions and AGN-ionised H\,\textsc{ii} regions, we overlaid the H\,\textsc{ii} region maps onto our FUV ellipses. For a precise geometric assessment, we employed the Shapely library in Python. Each FUV region was modelled as an ellipse, characterised by its semi-major axis, semi-minor axis, and position angle as determined by ProFound, while H\,\textsc{ii} AGN regions were represented as circles, defined by their RA, Dec, and measured areas from \citet{2023MNRAS.520.4902G}. We used the intersection method in Shapely to accurately determine whether the elliptical FUV regions and circular AGN regions shared any common area. Based on this analysis, we classified the regions into two categories: (i) FUV regions that overlap with AGN-ionised regions and (ii) FUV regions with no AGN overlap. The former were excluded from further analysis to ensure that our results focus solely on star formation unaffected by AGN contamination.

\subsection{Extinction correction using MUSE Data}
To accurately estimate the SFR in the star-forming regions using UV emission, it is crucial to consider both internal extinction within the galaxies and the extinction along the line of sight due to our Milky Way. We compiled the ratio of H$\alpha$ and H$\beta$ flux values ($F_{\mathrm{H}\alpha} / F_{\mathrm{H}\beta}$
) for the identified star-forming regions from \citet{2021yCat..36580188S}. The number and morphology of the H\,\textsc{ii} regions from \citet{2021yCat..36580188S} differ from the regions identified in our work. To begin, we selected the FUV regions within the
MUSE coverage using Topcat. Spatial correspondence was then established by transforming celestial
coordinates into pixel coordinates and checking which H\,\textsc{ii} regions fell within the elliptical boundaries
of each SF region. For SF regions containing multiple H\,\textsc{ii} regions, we calculated the mean $F_{\mathrm{H}\alpha} / F_{\mathrm{H}\beta}$ and its associated uncertainty. In cases where no H\,\textsc{ii} regions overlapped, the algorithm used the
10 closest H\,\textsc{ii} regions to compute these values. For the internal extinction correction, we adopted an intrinsic Balmer ratio of 2.86 for case B recombination at an electron temperature T$_e$ = 10,000 K and a density n$_e$ = 100 cm$^{-3}$ \citep{1989agna.book.....O}. Extinction in H${\alpha}$ is estimated as follows:,
\begin{equation}
A(H\alpha) = \frac{K_{H\alpha}}{-0.4 \times (K_{H\alpha} - K_{H\beta})} \times \log \left( \frac{F_{H\alpha} / F_{H\beta}}{2.86} \right)
\end{equation}

where K$_{{H{\alpha}}}$ = 2.53 and K$_{{H{\beta}}}$ = 3.61 are the extinction coefficients for the Galactic extinction curve from \citet{1989ApJ...345..245C}. The obtained A(H${\alpha}$) values were then converted to A$_{FUV}$ (extinction in the FUV wavelength) using the coefficients provided by \citet{1989ApJ...345..245C}. Additionally, we considered the line of sight extinction using the values of A$_V$ compiled for each galaxy from \citet{2011ApJ...737..103S}. Subsequently, we used the extinction-corrected magnitudes for further analysis.


\subsection{Estimation of $\Sigma_{SFR}$}
The SFR in each region is computed using the relation derived from \citet{2013AJ....146...46K}, as presented below:
\begin{equation}
\log(SFR_{FUV} \, (M_{\odot} \, \mathrm{yr}^{-1})) = 2.78 - 0.4 \, mag_{FUV} + 2 \log(D)
\end{equation}
Here, $mag_{FUV}$ denotes the extinction-corrected magnitude, and $D$ represents the distance to the galaxy in Mpc. The ProFound package identifies the total number of pixels containing 100$\%$ of the flux for each region. We then used the galaxy's distance in Mpc to convert the angular size of each star-forming region into a physical size in kpc$^2$. The $\Sigma_{SFR}$ for each region was then calculated by dividing the estimated SFR by its corresponding area and then multiplying by \emph{cos(i)}, where \emph{i} is the galaxy's inclination angle.

\subsection{Calculation of $\Sigma_{H_2}$}
In order to establish a correlation between the properties of the molecular gas and the corresponding UV emission, it is imperative to pinpoint the same star-forming regions obtained from UV in the ALMA CO(2-1) images. This involves overlaying the identified star-forming regions from UV images onto the ALMA images to track the CO(2-1) emission across the active star-forming regions of each galaxy captured in the corresponding UVIT images. This overlay process is facilitated by ProFound, allowing us to extract the line-integrated CO(2-1) intensity. We selected only those regions where the CO(2-1) signal-to-noise ratio (\emph{S/N}) $>$ 3, ensuring that our measurements are not dominated by noise. Subsequently, we computed $\Sigma_{H_2}$ using the method outlined in \citet{2021ApJS..257...43L}. 
\begin{equation}
    \Sigma_{H_2} \, (\mathrm{M_{\odot} \, pc^{-2}}) = \alpha_{CO} \times R_{21}^{-1} \times I_{\mathrm{CO(2-1)}} \, [\mathrm{K \, km \, s^{-1}}] \times \cos(i)
\end{equation}
where \emph{$\alpha_{\rm CO}$} is the CO(1--0) conversion factor, \emph{$R_{21}$} \citep[= 0.65;][]{2013AJ....146...19L,2021MNRAS.504.3221D} is the CO(2--1)-to-CO(1--0) line ratio in Kelvin, \emph{$i$} is the inclination of the galaxy, and \emph{$I_{\rm CO(2-1)}$} is the line-integrated CO(2--1) intensity in K km s$^{-1}$. The simplest approach is to assume a constant MilkyWay-like CO-to-H$_2$ conversion factor of $\alpha_{\rm CO} = 4.35 \, M_\odot \, (\mathrm{K \, km \, s^{-1} \, pc^2})^{-1}$ \citep{2013ARA&A..51..207B}. However, this factor is known to strongly depend on gas metallicity \citep{2017MNRAS.470.4750A}, and various parameterisations of this dependence have been proposed \citep{2013ARA&A..51..207B}. In this study, we utilise the metallicity-dependent relation derived by \citet{2017MNRAS.470.4750A}, which has also been adopted in recent studies by \citet{2020ApJ...901L...8S,2023MNRAS.519.1149B}, and is expressed as:

\begin{equation}
    \alpha_{\rm CO} = 4.35 \times \left(\frac{Z}{Z_\odot}\right)^{-1.6} \, M_\odot \, (\mathrm{K \, km \, s^{-1} \, pc^2})^{-1}.
\end{equation}

We calculated the mean gas-phase metallicities (\emph{Z/Z$_{\odot}$}) for each region based on the 12 + log(O/H) measurements from \citet{2023MNRAS.520.4902G}. These values were derived using the S calibration method \citep[][]{2016MNRAS.457.3678P}, which was applied to H\,\textsc{ii} regions identified in VLT/MUSE data (see \citet{2023MNRAS.520.4902G}.)

\section{Results}
\label{sec:results}
\subsection{Resolved Kennicutt-Schmidt Relation}

\begin{table*}[h]
\centering

\caption{Correlation coefficients and orthogonal distance regression parameters for the sample galaxies.}\label{table:fit}

\begin{tabular}{llllll}
\hline
Galaxy & Resolution  & \multicolumn{2}{c}{Correlation Coefficient} & \multicolumn{2}{c}{Orthogonal Distance Regression Parameters} \\
         & (pc)       & Pearson              & Spearman             & Slope(N)                    & Intercept(A)                      \\ \hline
NGC 1365 & 132        & 0.43                 & 0.26                 & 0.17 $\pm$ 0.09        & -1.57 $\pm$ 0.13          \\ \hline
NGC 1433 & 126        & 0.60                 & 0.47                 & 0.30 $\pm$ 0.09        & -2.15 $\pm$ 0.10          \\ \hline
NGC 1566 & 120        & 0.71                 & 0.74                 & 0.71 $\pm$ 0.04        & -2.74 $\pm$ 0.07          \\ \hline     
\end{tabular}
\end{table*}


In this section, we present the results of rK-S relation for each galaxy in our sample. Using FUV emission for $\Sigma_{SFR}$ and $^{12}$CO(2-1) emission for $\Sigma_{H_2}$, we conducted our analysis at spatial resolutions ranging from $\sim$ 120 pc to 132 pc in three nearby galaxies.

\begin{figure*}[t]
    \centering
    \label{fig-ksrelation}
    \includegraphics[width=1\linewidth]{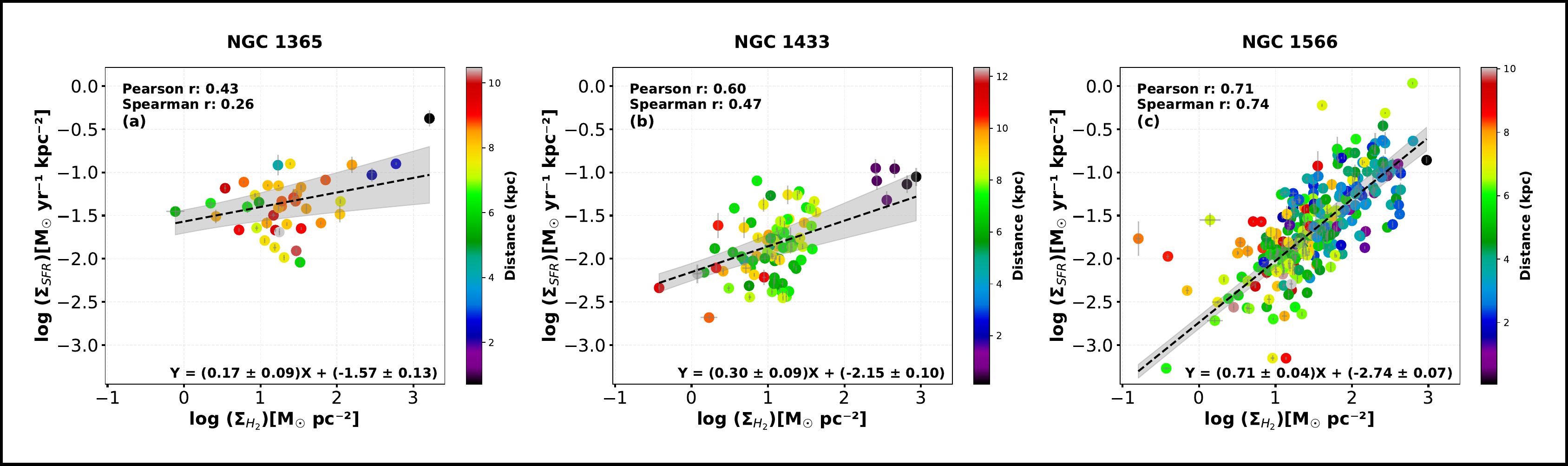}
    
    \caption{The figure illustrates the rK-S law for the sample galaxies, depicting $\Sigma_{SFR}$ as a function of $\Sigma_{H_2}$, accompanied by their respective uncertainties in estimation. Each data point is colour-coded according to the distance (r) between the corresponding region and the centre of the galaxy. The black dashed lines represent the best-fit relations for the data, determined using orthogonal distance regression. The equation of the best-fit line is provided within each plot. The grey shaded region indicates the 1-$\sigma$ uncertainty around the fit, which reflects the uncertainties in the fit parameters (slope and intercept).}
    
\end{figure*}

The rK-S relation for the sample galaxies is shown in Figure \ref{fig-ksrelation}. Significant variations in the rK-S relations among individual galaxies are illustrated in this figure. For example, at a resolved scale of 120 pc, the galaxy NGC 1566 exhibits a strong relation (Pearson correlation coefficient = 0.71 and Spearman correlation coefficient = 0.74, see Figure \ref{fig-ksrelation}(c)) with a slope of 0.71 $\pm$ 0.04. In contrast, the other two galaxies in the sample exhibit weaker correlations. NGC 1365, at a resolved scale of 132 pc, has a Pearson correlation coefficient of 0.43 and a Spearman correlation coefficient of 0.26 (see Figure \ref{fig-ksrelation}(a)), while NGC 1433, at a resolved scale of 126 pc, shows a Pearson correlation coefficient of 0.60 and a Spearman correlation coefficient of 0.47 (see Figure \ref{fig-ksrelation}(b)). We fitted the $\Sigma_{SFR}$ versus $\Sigma_{H_2}$ relation using an orthogonal distance regression \citep[ODR;\footnote{\url{https://docs.scipy.org/doc/scipy/reference/odr.html}}][]{2020NatMe..17..261V, 2020A&A...634A..24K} algorithm. The fitting parameters are summarized in Table \ref{table:fit}. Utilizing high-resolution UVIT and ALMA data, we identified rK-S relation indexes (N) ranging from $\sim$ 0.17 to 0.71. For the galaxy NGC 1566, the slope value 0.71 is nearly close to the values reported in the literature \citep[e.g.][]{2008AJ....136.2846B,2010ApJ...714L.118D,2012MNRAS.421.3127N,2014MNRAS.442.2208S,2015A&A...577A.135C}. It is important to note that the slope of the rK-S relation can vary significantly depending on the CO-to-H$_2$ conversion factor used, the resolution, and the specific fitting methodology employed \citep{2013AJ....146...19L,2021MNRAS.505L..46E,2021A&A...656A.133Q,2023A&A...671A...3J}. 

Our finding contrasts with the conclusion of \citet{2021A&A...650A.134P} who observed that the rK-S relation shows the lowest level of scatter at lower spatial resolutions. They argue that the rK-S relation exhibits greater consistency across different galaxies and environments, with similar slope and normalisation values, suggesting that the SFR follows the molecular gas distribution over short timescales, based on their identification of star-forming regions using H$\alpha$ emission. However, this conclusion may not apply to FUV emission, which traces star formation over timescales of $\sim$ 100 Myr. Additionally, the impact of resolution on the slope depends largely on how non-starforming resolution elements are treated during the fitting process, as noted by \citet{2021A&A...650A.134P}.

Moreover, defining a ``star-forming region'' becomes more nuanced when using FUV emission compared to H$\alpha$. H$\alpha$ traces the ionizing radiation from massive stars over short timescales \citep[$\sim$5--10 Myr;][]{2024ApJ...976...90T}, whereas FUV emission captures star formation activity over significantly longer timescales ($\sim$10--100 Myr; \citealt{2012ARA&A..50..531K}). Consequently, FUV observations represent the integrated light from a broader and more temporally diverse population of young stars, potentially encompassing multiple or sequential star-forming events. As a result, FUV-identified regions tend to be more extended and diffuse \citep{2023MNRAS.526.1512R}, with their morphology further affected by processes such as stellar migration and dust scattering \citep{2005ApJ...619L..67T}. These factors contribute to the heterogeneous sizes of the detected regions and represent an intrinsic limitation of UV-based studies, introducing a possible caveat in the interpretation of our results.


The weaker rK-S relations observed in NGC 1365 and NGC 1433 may be attributed to the insufficient presence of molecular gas in regions other than their centres. In contrast, NGC 1566 exhibits a more uniform distribution of molecular gas, both in the centre and the inner spiral arms, which facilitates efficient star formation. This uniform distribution is reflected in the strong rK-S relation observed in NGC 1566. A sub-linear slope in the rK-S relation suggests that star formation efficiency decreases with increasing $\Sigma_{H_2}$. This indicates that there may not be a direct, one-to-one correlation between CO emission and star formation, as noted by \citet{2014MNRAS.437L..61S}.
However, our findings reveal a crucial correlation between host-galaxy properties and the interplay of molecular gas and star formation tracers. Specifically, the relationship between molecular gas and SFR tracers on the K-S plane varies across galaxies, influenced by how molecular gas and star formation are spatially distributed - whether centrally concentrated or more extended - consistent with earlier studies \citep[e.g.][]{2008AJ....136.2846B,2021MNRAS.505L..46E,2023A&A...671A...3J}.

\subsection{Star formation Efficiency in the sample}
\begin{figure*}[t]
    \centering
    \includegraphics[width=0.75\linewidth]{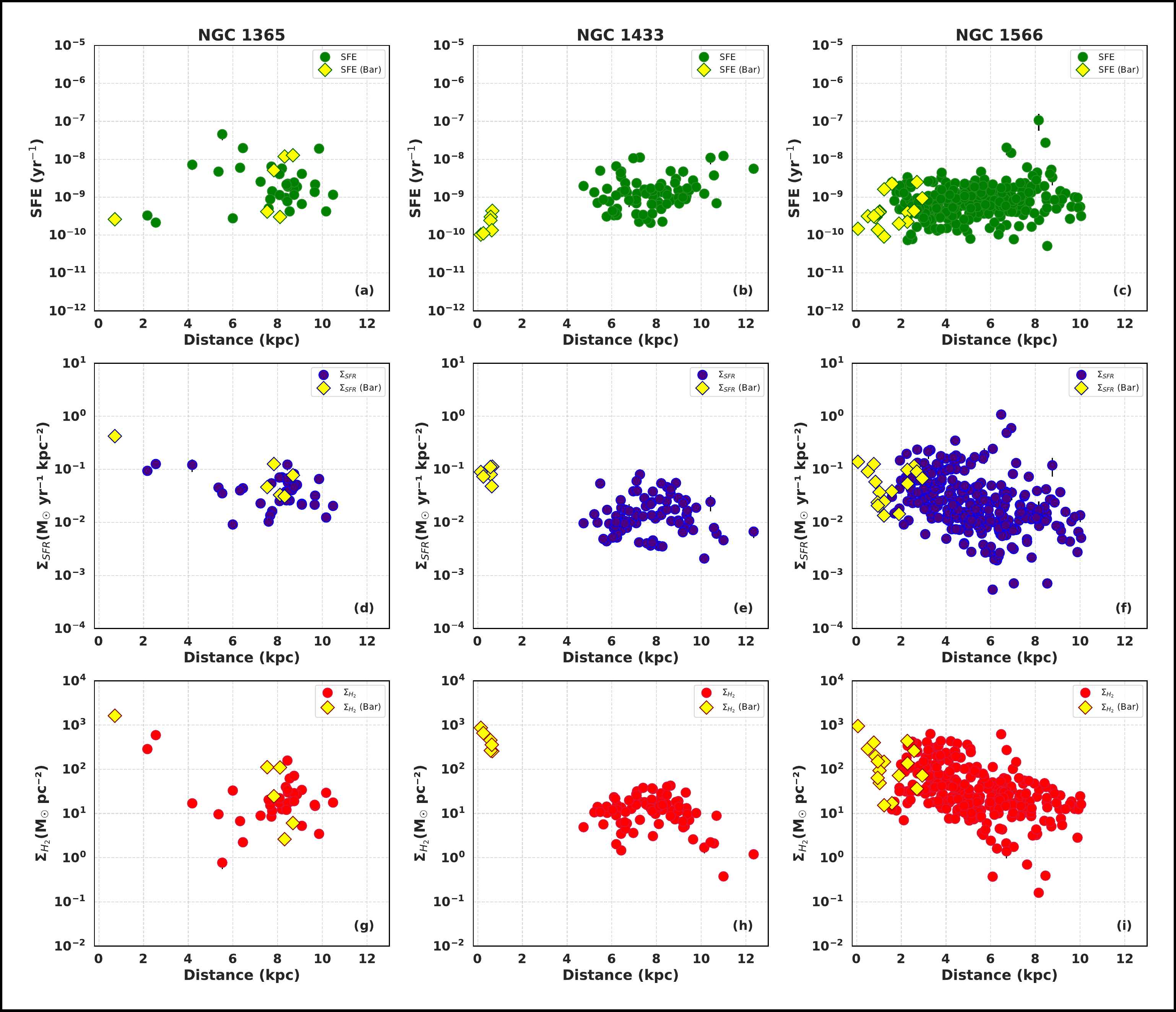}
    
    \caption{The figure shows the variation of $\Sigma_{H_2}$ (bottom row), $\Sigma_{SFR}$ (middle row), and SFE (top row), with radial distance in kpc on the X-axis and values on a logarithmic Y-axis, for the sample galaxies. The yellow diamonds represent regions within the bar of each galaxy, with the bar extent determined through isophotal analysis, as described in Section \ref{subsec:bar}.
    It is clear that, while $\Sigma_{H_2}$ and $\Sigma_{SFR}$ peak in the central regions, SFE remains notably low, suggesting limited SFE despite high molecular gas concentrations.}
    
    \label{fig-sfe}
\end{figure*}

Star formation efficiency (SFE) is a metric used to assess how efficiently gas within molecular clouds is converted into stars over a given time period \citep[e.g.][]{2008AJ....136.2782L,2010ApJ...721..383M}. It quantifies the rate of star formation relative to the amount of available gas \citep{2024MNRAS.528.4393G}.
We calculated the SFE of each region using the following definition adopted from \citet{2024arXiv240505364Q}:
\begin{equation}
    SFE \, (\mathrm{yr}^{-1}) = \frac{\Sigma_{SFR} \, (M_{\odot} \, \mathrm{yr}^{-1} \, \mathrm{kpc}^{-2})}{\Sigma_{H_2} \, (M_{\odot} \, \mathrm{kpc}^{-2})}
\end{equation}
The SFE with the radial distance of the galaxy from the centre is shown in Figure \ref{fig-sfe} (top panel). The SFE in the sample galaxies appears to be exceptionally low in their central regions, including the nuclear areas and the bar, with the nucleus potentially encompassing a portion of the bar regions. This could be linked to the channelling of gas towards the centre, which might suggest an increase in star formation; however, this accumulation can also lead to increased turbulence \citep{2021AJ....161..243S} and complex interactions \citep[e.g.][]{2013ApJ...769...82S}, ultimately diminishing SFE in these central areas. The influence of both AGN \citep{2019A&A...624A..81M} and the bar structure \citep{2020A&A...644A..38D} in our sample may significantly affect this process \citep{2017MNRAS.469.3722R}, although the precise contributions of each remain unclear from the analysis. The literature indicates that AGN and bars influence star formation efficiency (SFE) through distinct mechanisms, each leading to different outcomes \citep[e.g.,][]{2016A&A...595A..63V,2023MNRAS.524.3130S}. AGN can exert both positive and negative feedback, modifying the conditions for star formation in their surrounding environments \citep{2019ApJ...881..147S}. In contrast, bars primarily affect gas dynamics and increase gas density in the central regions of their host galaxies \citep{2024AAS...24330610M}.


\citet{2021ApJ...913..139G} also detected high-density regions with $\Sigma_{H_2} > 100$ M$_\odot$ pc$^{-2}$ but with markedly lower star formation efficiency (SFE $\sim 10^{-11}$ yr$^{-1}$) within the galaxy NGC1365, suggesting an inhibition of star formation in multiple dense gas regions. Conversely, in the low-density regime, they observed numerous areas displaying relatively elevated SFE ($\sim 10^{-8}$ yr$^{-1}$). These regions are distributed across both inner and outer regions, indicating significantly enhanced star formation.

\begin{figure*}[t]
    \centering
    \includegraphics[width=\linewidth]{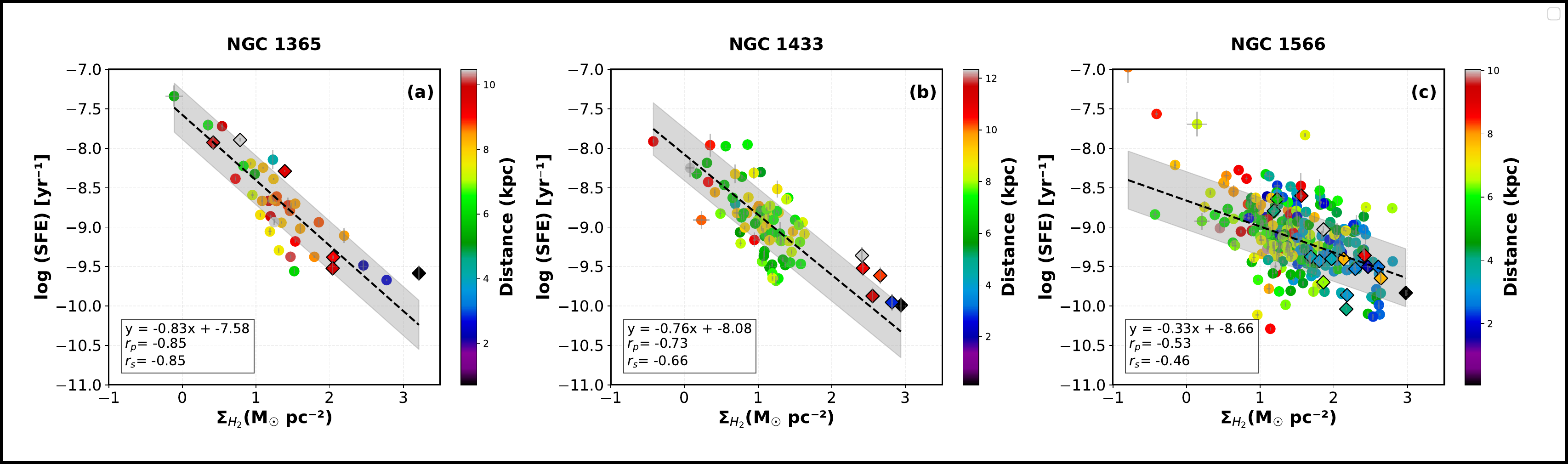}
    
    \caption{The figure shows the correlation between star formation efficiency (SFE) and molecular gas surface density ($\Sigma_{H_2}$) for the sample galaxies, with colour coding indicating the distance of the star-forming regions from the centre galaxy. The dotted line represents the linear regression best-fit line, while the correlation coefficients, denoted as r$_p$ (Pearson) and r$_s$ (Spearman), are provided within the plots. The shaded region represents the 1-$\sigma$ confidence interval for the fit. It is evident that higher $\Sigma_{H_2}$ leads to lower SFE in the sample galaxies. The diamond symbols represent regions within the bar of each galaxy, where the bar extent is determined through isophotal analysis, as described in Section \ref{subsec:bar}. Notably, the regions inside the bar follow the same trend as those outside it.} 
    
    \label{fig-sfe_sigma_mol}
\end{figure*}

We investigated the relationship of SFE with $\Sigma_{H_2}$ and $\Sigma_{SFR}$ to discern their respective contributions to SFE in these galaxies. Figure \ref{fig-sfe_sigma_mol} illustrates the correlation between SFE and $\Sigma_{H_2}$ exhibited by the sample galaxies. In our study, a strong negative correlation (Pearson's coefficient = -0.85, Spearman's coefficient = -0.85, see Figure \ref{fig-sfe_sigma_mol}(a)) is found in NGC 1365 and good correlation (Pearson's coefficient = -0.73, Spearman's coefficient = -0.66, see Figure \ref{fig-sfe_sigma_mol}(b)) in NGC 1433. In contrast, NGC 1566 shows a weaker (Pearson's coefficient = -0.53, Spearman's coefficient = -0.46, see Figure \ref{fig-sfe_sigma_mol}(c)) relationship, despite exhibiting a robust rK-S correlation. We also examined the relationship between the SFE and $\Sigma_{H_2}$ within the central 4 kpc region of NGC 1566, in order to compare it with the ALMA field of view in the other galaxies in our sample. The results showed improved correlation values (Pearson's coefficient = -0.68, Spearman's coefficient = -0.66), indicating that this relationship is more pronounced in the central regions of these galaxies than in the outer areas, including the spiral arms. Additionally, we explored the relation between SFE and $\Sigma_{SFR}$ in sample galaxies but could not observe any significant trend.

In their study, \citet{2020MNRAS.495.3840M} identified a negative correlation between SFE and the diffuse molecular gas fraction (f$_{dif}$) within the strongly barred galaxy NGC 1300. They observed that the molecular gas traced by $^{12}$CO(1-0) comprises two distinct components -- first one due to GMCs and an extended diffuse component distributed over scales larger than sub-kpc \citep{2020MNRAS.495.3840M}. This finding aligns with the notion that a higher abundance of diffuse molecular gas leads to a lower SFE. They noted a tighter relationship when utilizing FUV and 22$\mu$m as tracers for SFR compared to H$\alpha$ emission. The trends observed in Figure \ref{fig-sfe_sigma_mol}(a) and Figure \ref{fig-sfe_sigma_mol}(b) support their conclusions, albeit discussing the general molecular gas rather than its specific components.

The star formation process is inherently dynamic, with molecular clouds evolving from gas-rich, starless states to active star-forming regions before being dispersed by stellar feedback \citep{2017MNRAS.464.3536R,2020MNRAS.493.2872C}. As a result, molecular clouds do not follow a universal star formation relation at any given time, since their gas content and SFE vary across different evolutionary stages \citep{2020MNRAS.493.2872C}. This is evident when selecting star-forming regions based on molecular gas peaks (e.g., CO emission) versus star formation peaks (e.g., H$\alpha$ or UV emission). Studies have demonstrated that the gas-to-SFR ratio depends on this selection method, influencing the inferred timescales of the star formation process \citep{2019Natur.569..519K,2020MNRAS.493.2872C}

In this study, we aim to investigate star formation at spatial scales comparable to those of individual GMCs, using UV emission as the SFR tracer instead of H$\alpha$. Unlike H$\alpha$, which predominantly traces massive stars over short timescales \citep[$\sim$5--10 Myr;][]{2024ApJ...976...90T}, UV emission is sensitive to star formation over significantly longer periods \citep[$\sim$100 Myr;][]{2009ApJ...706..553B,2012AJ....144....3L}. Consequently, UV-based measurements tend to average over multiple evolutionary stages of the star formation process, thereby smoothing out temporal fluctuations in the SFR. As a result, the regions identified via UV emission may reflect the cumulative output of several star-forming events rather than isolated, individual GMCs. This temporal averaging can lead to systematically lower estimates of the SFE when compared to results derived from H$\alpha$, as UV traces both the early and more evolved phases of star formation activity \citep[e.g.,][]{2010ApJ...722.1699S}. Moreover, we are likely probing the later phases of the star formation cycle, when much of the surrounding molecular gas has already been dispersed. This methodological difference is crucial in differentiating our results from studies that use molecular gas-based region selection, which typically focus on earlier phases of the GMC lifecycle.

The spatial resolution of our study also impacts the interpretation of star formation scaling relations. At GMC scales, the observed correlation between gas and SFR is strongly influenced by the evolutionary stage of the regions being studied \citep{2019Natur.569..519K,2020MNRAS.493.2872C}. Given that UV emission integrates over longer timescales, it blends multiple evolutionary stages, making it more challenging to isolate active star-forming regions from those transitioning between phases. These timescale effects must be considered when comparing SFE measurements across different studies. While we recognize that FUV imaging does not offer the same spatial resolution as H$\alpha$, it
remains a robust and direct tracer of recent star formation, capturing light from young, massive stars.
Despite these limitations, our analysis makes use of the highest UV spatial resolution currently available,
allowing us to probe star formation down to physical scales of $\sim$ 100-200 pc. We believe this approach provides valuable
insights into the spatial distribution and efficiency of star formation, even when individual compact
regions are not fully resolved.

\subsection{SFE within the bar regions of the sample galaxies}
\label{subsec:bar}
\begin{figure}
\includegraphics[width=\columnwidth]{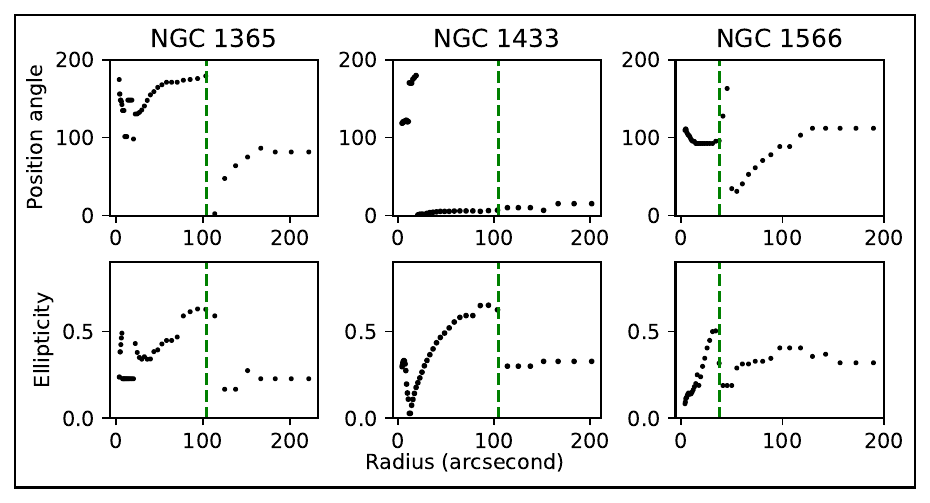}
\caption{The figure illustrates the variation in position angle and ellipticity derived from isophotal analysis of the sample galaxies using Spitzer/IRAC 3.6$\mu$m data. Abrupt changes in these parameters mark the radial extent of the bar in each galaxy.}
\label{fig:isophotal}
\end{figure}

Stellar bars are common morphological structures in the local Universe and are believed to play a crucial role in secular evolutionary processes. They efficiently redistribute gas, stars, and angular momentum within their host galaxies. Two fundamental quantities that characterise a bar are its length and strength. Various methods can estimate the characteristic size of a bar. By analyzing variations in brightness distributions, features such as bars and rings within galaxies can be identified \citep{2007ApJ...657..790M, 2024A&A...681A..35A}. For bar components, common approaches include measuring the maximum bar ellipticity \citep{2002ApJ...567...97L} and examining the radial variation of the position angle \citep{2005ApJ...626L..81E}. Isophotal analysis is an effective technique for studying various properties of galaxies at different surface brightness levels. This method provides valuable insights into the structure, morphology, and composition of galaxies \citep{2009ApJS..182..216K}.

\begin{figure*}
    \centering
    \includegraphics[width=0.82\linewidth]{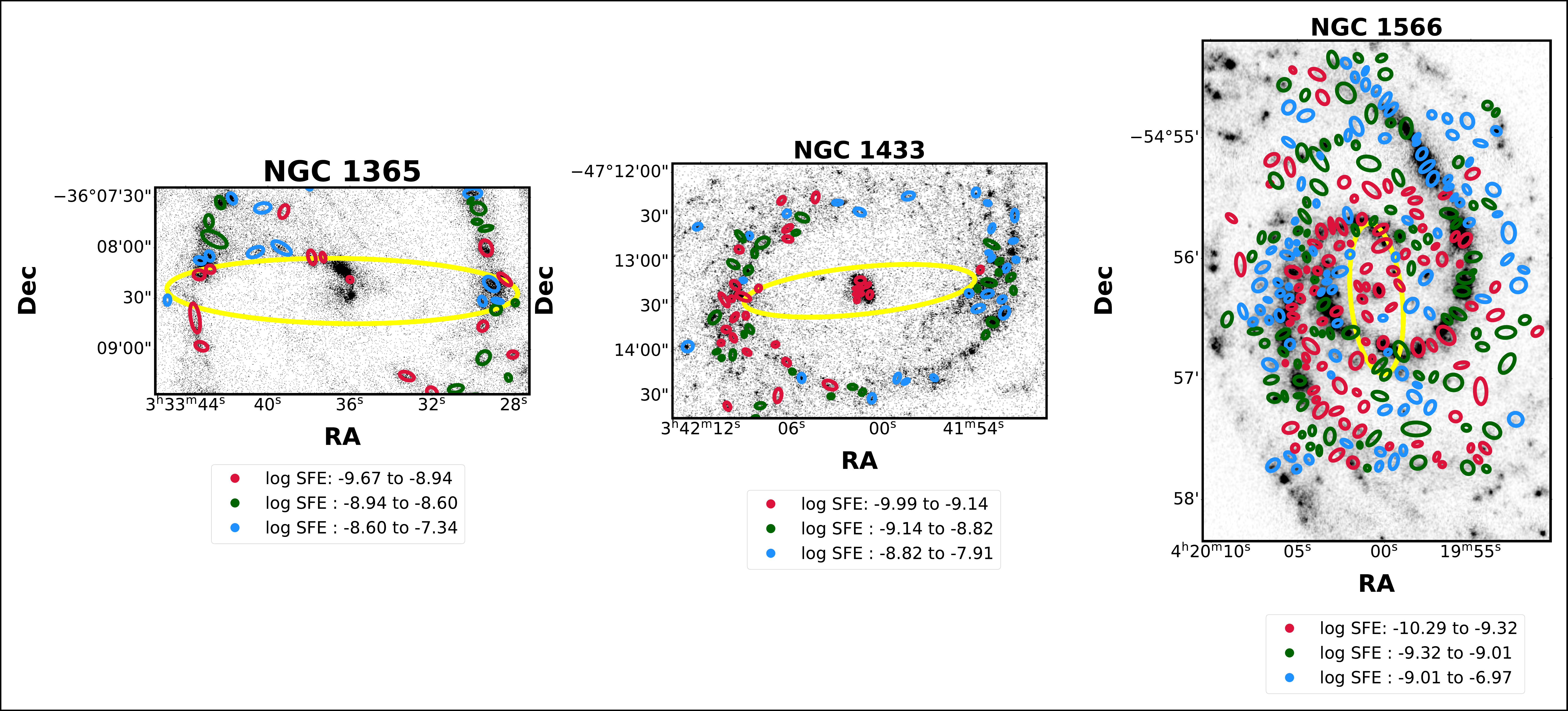}
    
    \caption{The figure displays the star-forming regions in the sample galaxies, colour-coded by their SFE. Red indicates lower SFE, green represents medium SFE, and blue signifies high SFE. The yellow ellipse marks the extent of the bar in each galaxy, determined through isophotal analysis of \emph{Spitzer}/IRAC 3.6 $\mu$m images. The figure clearly shows that the yellow ellipse encloses regions with lower SFE. For all the images North is up and East to the left.}   
    
    \label{fig-bar-sfe}
\end{figure*}

To determine the radial extent of the bar component in our sample galaxies, we performed surface photometry using \emph{Spitzer} (IRAC) 3.6$\mu$m images, by applying elliptical isophote analysis in Python\footnote{\url{https://photutils.readthedocs.io/en/stable/isophote.html}}, following the methodology described by \citet{1987MNRAS.226..747J}. During the fitting process, the centre, ellipticity, and position angle (measured counterclockwise from the west) of the isophotes were allowed to vary freely. The radial extent of the bar component in each galaxy was determined using isophotal parameters such as position angle and ellipticity. Abrupt changes in these parameters were used to identify the bar's radial extent in each galaxy (see Figure \ref{fig:isophotal}). Based on our analysis and observations of the IRAC images, we estimated the bar radius to be 10.81 $\pm$ 0.95 kpc for NGC 1365, 7.73 $\pm$ 0.77 kpc for NGC 1433, and 3.54 $\pm$ 0.34 kpc for NGC 1566. The bars are indicated in yellow coloured ellipses in Figure \ref{fig-bar-sfe}.

In top panel of Figure \ref{fig-sfe} the variation of SFE with galactocentric distance is plotted where yellow diamonds represents the regions within the bar of each galaxies. The SFE in the bars of the sample galaxies is notably low ($\sim$ 10$^{-8}$ to 10$^{-10}$), as evidenced by the majority of star-forming regions within the bar falling within this range. For galaxies NGC 1365 and NGC 1433, there seems to be less number of star-forming regions in the bar compared to the spiral arms (see left and middle panels of Figure \ref{fig-bar-sfe}), suggesting a lower rate of recent star formation in these regions. While NGC 1566 has more star-forming regions in the bar (right panel of Figure \ref{fig-bar-sfe}) compared to the other two galaxies in the sample, the efficiency remains low, similar to that observed in the other galaxies.

To determine whether the lower SFE in bar regions is an intrinsic property of the bar structure or a consequence of the typically higher $\Sigma_{H_2}$ found in the central regions (see the bottom panel of Figure \ref{fig-sfe}), we conducted an analysis incorporating $\Sigma_{H_2}$ as a covariate. Figure \ref{fig-sfe_sigma_mol} presents a scatter plot of log SFE as a function of $\Sigma_{H_2}$, with bar regions marked by diamonds. The results reveal a clear correlation between SFE and $\Sigma_{H_2}$, where both bar and non-bar regions follow a similar trend. This suggests that at a fixed $\Sigma_{H_2}$, the SFE does not show a statistically significant difference between bar and non-bar regions. Although our sample lacks non-barred galaxies for direct comparison, our findings suggest that molecular gas density primarily governs the observed trends in SFE rather than the presence of a bar itself. Supporting this, \citet{2012ApJ...758...73S}, using data from the xCOLD GASS survey, investigated the dependence of SFE on molecular gas content and environment, concluding that SFE is predominantly influenced by local gas conditions rather than structural features like bars.


\section{Discussion}
\label{sec:discussion}
The high-resolution FUV data from UVIT, combined with $^{12}$CO(2–1) data from ALMA, provide an excellent opportunity to explore spatially resolved star formation activity in the nuclear regions of three nearby galaxies. 
This study examines the relationship between star formation and molecular gas distribution in these environments. While various factors may influence the observed trends, our analysis focuses on the connection between molecular gas and star formation without isolating the impact of specific mechanisms. These results offer valuable insights into the processes governing star formation in the central regions of galaxies.

The analysis performed for each galaxy revealed a diverse range of behaviours concerning the rK-S relations. Interestingly, no "single" molecular star formation relation was observed, despite the fact that the investigated galaxies belong to the same subclass of objects, namely nearby barred spiral galaxies hosting AGN. It's noteworthy that all derived quantities were subjected to the same methodology and achieved at comparable spatial resolutions. The key finding indicates that each galaxy exhibits its own unique SF relation, characterised by distinct power-law indices (N) and correlation coefficients, observed at spatial scales of approximately 120 - 132 pc.

Numerous studies have consistently reported a moderate to strong K–S relation between $\Sigma_{SFR}$ and $\Sigma_{H_2}$ on kpc to sub-kpc scales, with the slope N typically ranging from 0.5 to 1.4 \citep[e.g.,][]{2002ApJ...569..157W,2008AJ....136.2846B,2011ApJ...730L..13B,2011AJ....142...37S,2012ApJ...745..190L,2013ApJ...772L..13M,2017ApJ...849...26U,2018ApJ...863L..21K,2018ApJ...853..149S,2019ApJ...872...16D, 2021A&A...650A.134P,2022A&A...659A.102S,2023ApJ...945L..19S,2023ApJ...943....7M,2024MNRAS.535.2538Z,2025ApJ...979..217S}.  Our analysis supports these findings for NGC 1566 at a spatial scale of 120 pc. However, for NGC 1365 and NGC 1433, at spatial scales of 126 pc and 132 pc, respectively, the relationship appears more complex. The lower slope values observed in these galaxies are likely driven by the significant differences in the number of identified star-forming regions, with NGC 1566 containing a much larger sample (691 regions) compared to NGC 1365 (147 regions) and NGC 1433 (108 regions). The limited number of star-forming regions in the central regions of NGC 1365 and NGC 1433 likely has a more substantial impact on the observed variations in slope.

A direct comparison with \citet{2021A&A...650A.134P} is essential, as they also examined the rK-S relation for the galaxies in this study. Analyzing 18 star-forming galaxies at a 100 pc resolution, they found that the shape of the rK-S relation remains consistent across different galaxies and environments, including centres, rings, bars, spiral arms, and discs \citep{2021A&A...656A.133Q}. They argued that this uniformity reflects the direct link between molecular gas and star formation, as SFR closely follows the molecular gas distribution on short timescales \citep[H$\alpha$ emission traces star formation over $\sim$10 Myr;][]{2012AJ....144....3L}. For NGC 1566, they reported a strong rK-S relation with a near-linear slope of 1.04 $\pm$ 0.02, which is steeper than our slope value of 0.71 $\pm$ 0.04, but still within a similar range, making the results broadly consistent. In NGC 1365 and NGC 1433, they reported lower slopes of 0.94 $\pm$ 0.01 and 0.70 $\pm$ 0.02, respectively, which are generally higher than the slopes we obtained but still broadly similar to our results. These differences likely arise from variations in the SFR tracer or the specific regions analyzed.
A key distinction of our study is the use of UV emission as an SFR tracer, which captures star formation over longer timescales ($\sim$100 Myr) compared to H$\alpha$ ($\sim$10 Myr) used in \citet{2021A&A...650A.134P}. This difference may explain the shallower slopes in our analysis, as UV emission includes both ongoing and older star formation, potentially smoothing out local variations in the rK-S relation. By employing a different tracer, our study provides new insights into star formation across different environments, particularly in dynamically complex galaxies hosting bars and AGN. These results emphasise the importance of multi-wavelength studies in understanding star formation processes at various evolutionary stages.

This complexity may be further exemplified by previous studies \citep[][]{2010ApJ...722.1699S,2014MNRAS.439.3239K,2015A&A...578A...8B,2019ApJ...870...79S} showing that lower resolved scales can weaken the correlation between molecular gas and SFR. In NGC 1365 and NGC 1433, for example, the spatial overlap between these components is less consistent, leading to a weaker correlation \citep{2022ApJ...927....9P}. This mismatch introduces significant scatter in the rK-S relation at resolved scales of 126 pc and 132 pc, likely due to incomplete sampling of different phases of the gas and star formation cycles \citep[e.g.,][]{2014MNRAS.439.3239K}. The timescales associated with our SFR tracer further contribute to this complexity. Since FUV emission traces star formation over the past $\sim$100 Myr \citep{2011A&A...533A..19B,2012ARA&A..50..531K}, it may not capture the most recent star-forming events that are more closely linked to the current molecular gas distribution. This temporal offset between molecular gas and star formation can lead to increased scatter in spatially resolved studies \citep[e.g.][]{2010ApJ...722.1699S,2019Natur.569..519K}, particularly at small scales where individual regions may be at different evolutionary stages. \citep{2014MNRAS.439.3239K,2018MNRAS.479.1866K,2020MNRAS.493.2872C}


The slope (N) of the molecular rK-S relation remains a topic of debate \citep{2019ApJ...870...79S}. A correlation between SFR and molecular gas surface densities is expected, as both trace dense gas in the interstellar medium. In normal star-forming, non-starburst galaxies, this correlation is often found to be nearly linear \citep{2019ApJ...870...79S}, as we observed in NGC 1566. However, the relationship is not always linear, particularly in extreme environments such as starburst galaxies \citep{2010MNRAS.407.2091G,2015ApJ...800...20G} and galactic centres \citep{2013AJ....146...19L}.  While factors such as enhanced molecular gas depletion in the inner regions of galaxies due to processes like outflows, starburst activity, and stellar feedback can contribute to variations in the slope of the rK-S relation \citep{2019Natur.569..519K}, we find that the primary driver of the observed differences in N between NGC 1566 and the other two galaxies is the significantly lower number of identified star-forming regions in NGC 1365 and NGC 1433. Given this difference in sample size, statistical effects are likely the dominant factor influencing the measured slopes. However, physical processes, including AGN-driven feedback in NGC 1365 \citep{2024ApJ...974...36K} and NGC 1433 \citep{2013A&A...558A.124C}, could still play a secondary role in shaping the observed variations.

Additionally, the structural characteristics found in the central regions of galaxies may also impact the star formation law \citep[e.g.][]{2024MNRAS.532.4217S}, as demonstrated by the cases of NGC 1365 and NGC 1433. \citet{2021ApJ...913..139G} found that in the nuclear region of NGC 1365 at a resolution of 180 pc, the correlation coefficients were 0.67 (Pearson) and 0.72 (Spearman), with super-linear slopes of 1.67 $\pm$  0.10 ($\Sigma_{H_2}$ = 0 to 3.0 $M_{\odot}$ pc$^{-2}$) and 1.96 $\pm$  0.14 ($\Sigma_{H_2}$ = 1 to 3.0 $M_{\odot}$ pc$^{-2}$). Their analysis, which combined ALMA band 3 mapping and VLT/MUSE data, spanned the central 5.4 kpc of NGC 1365, offering valuable insights into the molecular and ionised gas distributions. In NGC 1365, the presence of a prominent central molecular zone (CMZ) is widely attributed to gas inflows driven by the highly non-axisymmetric, tumbling gravitational potential of stellar bars \citep{2024arXiv240319843S}. Although the exact physical mechanisms behind CMZ formation remain under active investigation, variations in molecular gas properties within the CMZ appear linked to inter-cloud motions in converging large-scale flows \citep{2023ApJ...944L..15S}. The SFR in the CMZ of NGC 1365 appears to be high, likely due to the inward transport of molecular gas by the stellar bar or spiral arms to the galaxy's centre \citep{2023ASPC..534...83H}. Additionally, \citet{2023ApJ...944L..14W} identified $\sim$ 30 young ($\leq$10 Myr), massive (M$_*$ $\geq$ 10$^6$ M$_\odot$) stellar clusters in NGC 1365's star-bursting centre using JWST near-IR imaging. A circum nuclear disk (CND) with a diameter $\leq$ 1 kpc has also been reported in NGC 1365, characterised by a lack of star formation. This CND may resemble the smooth molecular gas disks seen in early-type galaxies, where shear or tidal forces inhibit cloud collapse \citep{2023ApJ...944L..15S,2022MNRAS.512.1522D}. These structural features in NGC 1365 likely contribute to the observed weak rK-S relation in the galaxy as shown in Figure \ref{fig-ksrelation}(a).

In the case of NGC 1433, the presence of a nuclear ring \citep{2013A&A...558A.124C} within its nuclear region signals a site of intense starburst activity, as depicted in Figure \ref{fig-ksrelation}(b). \citet{2014A&A...567A.119S} propose that the bar of NGC 1433 efficiently channels gas towards the central region, where it accumulates in the nuclear ring, providing fuel for the intense star formation observed there. This mechanism may also explain the lack of dense molecular gas in the inner spiral arms of the galaxy and the absence of bright UV star-forming regions in the bar of NGC 1433. In this galaxy, bars appear to suppress star formation while inducing bursts of activity in the central regions. \citet{2023A&A...671A...8D} also identified the presence of an old bar in this galaxy, with an estimated age of approximately 7.5 billion years. \citet{2023A&A...671A...3J} reported variations in rK-S slopes between 0.69 and 1.40 in individual galaxies, with SFE differing by a factor of $\sim$ 4 between them. These studies underscore the variability in slope values, especially in regions with dense molecular gas, which is evident in the centres of NGC 1365 and NGC 1433. These findings highlight the role of environmental factors in contributing to the lower slopes observed.

In the case of NGC 1566, molecular gas appears to be more abundant in the central regions and along the spiral arms, leading to higher $\Sigma_{SFR}$ in these regions where dense gas is prevalent, thereby resulting in a favorable rK-S relation compared to the other galaxies in the sample. Despite being reported as not a starburst galaxy \citep{2004ApJ...606..271G}, the central region of NGC 1566 seems to have experienced a quiescent phase, with star formation activity only recently reinitiated. A plausible explanation could be the bar-driven secular evolution, which facilitated star formation in the centre by supplying gas to the innermost region \citep{2023A&A...673A.147P}. The density of star formation near the centre of NGC 1566 appears to be lower compared to the other galaxies in the sample, suggesting a weaker gas flow towards the centre. This results in the concentration of gas in the bar and spiral arms, promoting more uniform star formation throughout the galaxy and contributing to a moderately good rK-S relation, as shown in Figure \ref{fig-ksrelation}(c). In NGC 1566, we also observe significant scatter in the rK-S relation within the low-density molecular gas regime. A possible explanation for this phenomenon could be the substantial presence of atomic HI gas within this regime, which may result from the low conversion efficiency from the atomic to molecular phase, as suggested by prior studies \citep[e.g.][]{2021ApJ...913..139G}.

Recent or ongoing star formation on scales of 0.1-1 kpc around the nucleus is observed in all types of AGN, in contrast to quiescent galaxies \citep[e.g.][]{2022A&A...667A.145M,2024A&A...683L...8A}. However, the impact of AGN outflows on star formation - whether they quench or initiate it- remains debated \citep[e.g.][]{2014A&A...562A..21C,2017MNRAS.468.4956Z,2020MNRAS.497.3273F,2024Natur.630...54B}. It is likely that outflows can produce both effects. Strong star formation, characterised by young, bright, hot stars, can expel enough material from their gas clouds to trigger a series of accretion events onto the AGN \citep{2014A&A...567A.119S}.

On the other hand, the bar quenching hypothesis, supported by numerous recent observations \citep[e.g.][]{2024A&A...681A...7R,2024A&A...687A.255S} and simulations \citep[e.g.][]{2017MNRAS.465.3729S}, proposes that bars can instigate starburst events to fuel an AGN, accompanied by a rapid depletion of gas. This scenario potentially explains why some studies \citep[e.g.][]{2011MNRAS.416.2182E,2017MNRAS.469.3722R,2020MNRAS.499.1406L} indicate that barred galaxies exhibit higher SFR than unbarred counterparts, particularly in the central and circumnuclear regions \citep{2024arXiv240411656S}. This mechanism is notably evident in the galaxies NGC 1365 and NGC 1433 within our sample. The directed flow of cold gas towards the inner regions of these galaxies may contribute to the weakened relationship between molecular gas and star formation, as observed in the poor rK-S relation.

Recent discussions have questioned whether the star formation efficiency (SFE) varies across different galactic structures, particularly examining whether the SFE in the bar region is lower compared to other regions \citep{2023ApJ...943....7M}. As depicted in Figure \ref{fig-bar-sfe}, it becomes apparent that the star formation efficiency (SFE) within the bar regions of the sample galaxies remains notably lower in comparison to regions within their spiral arms. This result is consistent with previous literature \citep[e.g.][]{2023IAUS..373..207M}. The median SFE in the bars of our sample galaxies was found to be 0.95 Gyr$^{-1}$ for NGC 1365, 0.29 Gyr$^{-1}$ for NGC 1433, and 0.60 Gyr$^{-1}$ for NGC 1566. In comparison, \citet{2008AJ....136.2846B} investigated the K-S relation using CO data from the HERA CO Line Extragalactic Survey \citep[HERACLES;][]{2009AJ....137.4670L}, GALEX FUV, and \emph{Spitzer} 24 $\mu$m imaging across thousands of positions in 30 nearby disk galaxies at a 1 kpc resolution, reporting a median SFE of 0.43 Gyr$^{-1}$ within the bars. Similarly, \citet{2021A&A...654A.135D} measured SFE in the bars of 12 strongly barred galaxies, including NGC 5850 (0.25 Gyr$^{-1}$), NGC 4548 (0.24 Gyr$^{-1}$), and NGC 4535 (0.23 Gyr$^{-1}$), which were found to host H\,\textsc{ii} regions in the central parts of their bars. One potential explanation for the low SFE observed in barred galaxies is that the GMCs within these bars may be gravitationally unbound, as suggested by \citet{2013ApJ...779...45M,2013MNRAS.429.2175N}. However, \citet{2021A&A...654A.135D} argue that this phenomenon does not apply universally to all galaxies with strong bars, indicating that the relationship between bar strength and GMC stability may vary among different galactic environments.

\citet{2022A&A...663A..61P} investigated the rK-S relation in 18 galaxies from the PHANGS sample at a resolution of 150 pc, examining potential variations across different galactic environments. They found significant ($>$1$\sigma$) differences in the coefficients describing the rK-S relation, suggesting that additional physical processes influence star formation regulation. In particular, they reported lower SFE in bars, driven by radial and turbulent gas motions.
Similarly, \citet{2019PASJ...71S..15M} observed radial variations in SFE by a factor of 2–3 in individual galaxies from the CO Multi-line Imaging of Nearby Galaxies (COMING) project \citep{2019PASJ...71S..14S}. These results support the idea that local conditions, such as gas dynamics in bars and central regions, can lead to variations in the rK-S relation. 
\citet{2021A&A...656A.133Q} also analyzed 74 PHANGS galaxies, measuring $\Sigma_{H_2}$, $\Sigma_{SFR}$, and depletion times across galactic environments at a fixed spatial scale of $\sim$1.5 kpc. While they found a strong global correlation between $\Sigma_{H_2}$ and $\Sigma_{SFR}$ (with a slope of N = 0.97), they reported only minor variations across different environments. However, they noted systematically shorter depletion times in galaxy centres ($\tau_{dep}$ = 1.2 Gyr) and longer depletion times in bars ($\tau_{dep}$ = 2.1 Gyr). Our results are consistent with these trends, as we also find lower SFE in bars, supporting the idea that while bars can transport gas toward the center, they do not necessarily enhance SFE.  By using UV emission as an SFR tracer, our work extends these previous studies by providing a longer-timescale perspective on star formation.

Additionally, SFE within bars have been extensively studied in literature \citep[e.g.,][]{1996MNRAS.283..251K,2016PASJ...68...89M,2019PASJ...71S..13Y}. Studies indicate that star formation can be suppressed in bars, even in the presence of abundant molecular gas \citep{2020MNRAS.495.3840M}. Additionally, SFEs in spiral arms tend to be higher than those in bars \citep{2010ApJ...721..383M,2014PASJ...66...46H}. Our findings are consistent with these previous studies, reinforcing the idea that while bars can trigger nuclear starbursts and contribute to bulge formation, they do not significantly increase SFE compared to the rest of the galaxy population. Barred galaxies serve as examples wherein the star formation process can be locally influenced, as inward gas flows lead to heightened central gas concentrations and SFRs \citep{2012MNRAS.424.2180M}. Conversely, certain galaxies exhibit resistance to star formation, even in the presence of cold gas. The lack of recent star formation in the bars or the presence of old bars in the galaxies NGC 1365 and NGC 1433 may also contribute to the reduction in SFE in these galaxies. \citet{2016A&A...586A..45S} suggest that if old stars contribute to dust heating \citep[e.g.][]{2011AJ....142..111B,2019A&A...624A..80N}, it could lead to a decrease in SFE.

Indeed, numerous resolved studies emphasise that conditions within the ISM at galactic centres differ significantly from other regions, suggesting that the SFE is influenced by high stellar densities and/or the presence of AGNs \citep{2011A&A...534A..12B}. \citet{2016A&A...586A..45S} identified that in Centaurus A, the AGN jet or wind may lead to very inefficient star-forming molecular gas regions. Additionally, \citet{2005ApJ...629..680H} propose that AGNs suppress star formation in a sample of low-redshift quasars by injecting energy and momentum into the ISM through AGN feedback. Analysis of ALMA images of our sample galaxies reveals small outflows of gas from the AGN in the bar regions. These outflows may remove gas from the galaxy, which can reduce the potential for future star formation and result in a lower SFE. Supporting this, \citet{2017MNRAS.468.4205K} also propose that the observed outflow in molecular gas images may be depleting the cold molecular gas in the host galaxy. Furthermore, \citet{2017MNRAS.468.4205K} discuss that if the outflows lack the power to expel the gas, the intense ionizing radiation from AGNs might alter the state of the gas by either breaking down the molecules or ionizing them. An examination of the star formation history of the non-AGN sample might provide insights into these processes.

\section{Conclusion}
\label{sec:conclusion}
We investigated the resolved scale Kennicutt-Schmidt relation within the central regions of three nearby barred spiral galaxies hosting AGN, at spatial scales ranging from $\sim$ 120 to $\sim$ 132 pc. Below, we summarise the key conclusions drawn from our findings.
\begin{itemize}
\item We identified distinct rK-S relations unique to each galaxy, with slopes ranging from $\sim$ 0.17 to $\sim$ 0.71, indicating a sub-linear trend across the sample.
\item Each galaxy exhibits its own rK-S relation characterised by varying power-law indices and correlation strengths. These differences likely stem from the combined effects of the bar and AGN, although we could not fully separate their individual influences in our analysis.
\item NGC 1566 demonstrates a robust rK-S relation, with abundant molecular gas present in its central regions and spiral arms. The slope of the rK-S relation for NGC 1566 aligns well with values reported in existing literature.
\item In contrast, NGC 1365 and NGC 1433 display weaker correlations, likely attributed to the central concentration of molecular gas and distinct star formation patterns within their bars and nuclear regions. This complexity highlights the challenge of assuming a "single" rK-S relation across similar types of galaxies at these spatial scales.
\item We identified a break in the rK-S relation for NGC 1433 and NGC 1365 at spatial scales of $\sim$ 126 and $\sim$ 132 parsecs, respectively. We suggest that this break may be attributed to several factors:\\
i) The starburst nature of these galaxies.\\
ii) Structural features reported in NGC 1365 and NGC 1433.\\
iii) The directed flow of cold gas towards the centre of these galaxies as indicated by our analysis, potentially due to either a series of accretion events towards the AGN or bar quenching.\\
However, these findings emphasise the significance of environmental factors in shaping the rK-S relation.
\item Our study supports the notion that the rK-S relation is not always linear in extreme environments such as starburst galaxies and galactic centres. These conditions can significantly influence star formation efficiency (SFE), consequently modifying the slope of the relation.
\item We determined the median star formation efficiency in the bars of our sample galaxies to be 0.95 Gyr$^{-1}$ for NGC 1365, 0.29 Gyr$^{-1}$ for NGC 1433, and 0.60 Gyr$^{-1}$ for NGC 1566. These low SFE values are consistent with existing literature, which suggests that while bars can trigger nuclear starbursts and contribute to bulge formation, they do not significantly enhance SFE.
\item ALMA images of our sample galaxies reveal small outflows from AGN in the bar regions, which may also expel gas from the inner regions of the galaxies. This process leads to reduced SFE, which can be interpreted as a sign of negative feedback from AGN, as reported in the literature.
\item A negative correlation is observed between SFE and $\Sigma_{H_2}$ in the sample, indicating that higher molecular gas densities may result in lower SFE in the central regions of these galaxies.
\item The correlation between SFE and $\Sigma_{H_2}$ shows similar trends in both bar and non-bar regions, indicating that $\Sigma_{H_2}$ is the primary factor influencing SFE, rather than the presence of a bar in the sample.
\end{itemize}
In summary, our findings underscore the complex interplay between molecular gas and star formation mechanisms in the central regions of barred spiral galaxies hosting AGN, providing valuable insights into the intricate dynamics that shape their evolution.

\paragraph{Acknowledgement}
This work utilises data from the UVIT, which is part of the AstroSat mission by ISRO and is archived at the Indian Space Science Data Centre (ISSDC). The authors extend their sincere gratitude to all individuals involved in the various teams for their invaluable support, ranging from the early stages of design to launch and observations in orbit. This paper makes use of the ALMA data. ALMA is a partnership of ESO (representing its member states), NSF (USA) and NINS (Japan), together with NRC (Canada), NSTC and ASIAA (Taiwan), and KASI (Republic of Korea), in cooperation with the Republic of Chile. The Joint ALMA Observatory is operated by ESO, AUI/NRAO and NAOJ. Additionally, we acknowledge the utilization of data from PHANGS-ALMA and PHANGS-JWST. Furthermore, we express our appreciation to the NASA/IPAC Extragalactic Database (NED), supported by NASA and operated by the California Institute of Technology. We also acknowledge the use of NASA’s Astrophysics Data System Bibliographic Services. The authors extend their gratitude to the Centre for Research at CHRIST (Deemed to be University) for their unwavering support throughout this endeavour.

\paragraph{Data Availability Statement}
The UVIT observations utilised in this study are available in the AstroSat/ISSDC archive (\url{https://astrobrowse.issdc.gov.in/astro_archive/archive/Home.jsp}). Similarly, the PHANGS/ALMA data can be accessed at the Canadian Astronomy Data Centre (CADC) archive (\url{https://www.canfar.net/storage/list/phangs/RELEASES/PHANGS-ALMA/}). Upon reasonable request, the data depicted in each figure will be shared by contacting the corresponding author.    
\bibliographystyle{apalike}
\bibliography{example.bib}

\end{document}